\documentclass[twocolumn, pra, aps, superscriptaddress,longbibliography,showpacs,amsmath,amssymb,floatfix]{revtex4-1}
\usepackage{graphicx}
\usepackage{amssymb}
\usepackage{amsmath}
\usepackage{epsfig}
\usepackage{color}
\usepackage{tabu}
\usepackage{mathtools}
\usepackage[colorlinks,linkcolor=blue,anchorcolor=blue,citecolor=blue,urlcolor=blue]{hyperref}
\usepackage{physics}
\usepackage{float}
\usepackage{diagbox}

\begin{document}

\title{Anomalous isothermal compressibility in spin-orbit coupled degenerate Fermi gases}
\author{Cheng-Gong Liang}
\affiliation{Institute of Theoretical Physics, Shanxi University, Taiyuan, Shanxi 030006, P. R. China}
\affiliation{CA Key Laboratory of Quantum Information, University of Science and Technology of China, Hefei 230026, China}
\author{Yue-Xin Huang}
\affiliation{CA Key Laboratory of Quantum Information, University of Science and Technology of China, Hefei 230026, China}
\author{Fei-Hong Liu}
\affiliation{CA Key Laboratory of Quantum Information, University of Science and Technology of China, Hefei 230026, China}
\author{Yunbo Zhang}
\email{ybzhang@sxu.edu.cn}
\affiliation{Institute of Theoretical Physics, Shanxi University, Taiyuan, Shanxi 030006, P. R. China}
\author{Guang-Can Guo}
\affiliation{CA Key Laboratory of Quantum Information, University of Science and Technology of China, Hefei 230026, China}
\affiliation{Synergetic Innovation Center of Quantum Information and Quantum Physics, University of Science and Technology of China, Hefei, 230026, P.R. China}
\affiliation{CAS Center For Excellence in Quantum Information and Quantum Physics}
\author{Ming Gong}
\email{gongm@ustc.edu.cn}
\affiliation{CA Key Laboratory of Quantum Information, University of Science and Technology of China, Hefei 230026, China}
\affiliation{Synergetic Innovation Center of Quantum Information and Quantum Physics, University of Science and Technology of China, Hefei, 230026, P.R. China}
\affiliation{CAS Center For Excellence in Quantum Information and Quantum Physics}
\date{\today }

\begin{abstract}
The spin-orbit coupling (SOC) in degenerate Fermi gases can fundamentally change the fate of $s$-wave superfluids with strong Zeeman field and 
give rise to topological superfluids and associated Majorana zero modes. It also dramatically changes the thermodynamic properties of
the superfluids. Here we report the anomalous isothermal compressibility $\kappa_T$ in this superfluids with both SOC and Zeeman 
field. We formulate this quantity from the Gibbs-Duhem equation and show that the contribution of $\kappa_T$ comes from the explicit contribution of 
chemical potential and implicit contribution of order parameter. In the Bardeen-Cooper-Schrieffer (BCS) limit, this compressibility is 
determined by the density of state near the Fermi surface; while in the Bose Einstein condensate (BEC) regime it is determined by the scattering 
length. Between these two limits, we find that the anomalous peaks can only be found in the gapless Weyl phase regime. This anomalous behavior can be regarded 
as a remanent effect of phase separation. The similar physics can also be found in the lattice model away from half filling. These predictions can be measured 
from the anomalous response of sound velocity and fluctuation of carrier density.
\end{abstract}
\maketitle

The spin-orbit coupling (SOC) can modify the single-particle band structure \cite{sinova2004universal}, thus fundamentally change the 
behavior of ultracold atoms in the degenerate regime. In bosonic gases, the ground state can carry a 
finite momentum \cite{lin2011spin, galitski_spin-orbit_2013, stuhl_visualizing_2015}, giving rise to either plane wave phase or striped phase, depending on the interactions \cite{wang2010spin, ozawa2012ground, ozawa2012stability, cheng2018symmetry, sun2015tunneling, ho2011bose}.
The transition between these two phases can be described by Dicke model \cite{hamner2014dicke, zhang2013tunable, dicke1954coherence}.
Recently, this platform is used to search the supersolid phases\cite{leonard2017supersolid, li2017stripe, liao2018searching}. 
It can also be used to study the universal scaling of defects described by Kibble-Zurek mechanism during quench dynamics \cite{wu2017kibble, 
zurek_dynamics_2005}. The physics in Fermi gases are totally different due to the exclusive principle. 
The direct coupling between spin and momentum can make the spin polarization to be momentum dependent, thus when
an energy gap is opened by a Zeeman field, pairing is still allowed in the same band with $s$-wave 
interaction \cite{gong2011bcs, gong2012searching, chen2012bcs, iskin2011stability, yi2011phase, hu2011probing,
jiang2011rashba, yu2011spin, zhou2012opposite}. In case of inversion symmetry breaking, this system can support finite-momentum pairing
phases\cite{zheng2013route, wang2017fulde, he2018realizing, qu2013topological, zhang2013topological, zheng2015floquet, zheng2014fflo}. 
This mechanism was used in experiments for searching of topological phases and Majorana zero modes \cite{mourik2012signatures, 
deng2012anomalous, xu2015experimental, nadj2014observation, zhang2018quantized, das_zero-bias_2012, sun2016majorana, he2017chiral}. 
While the topological phases are widely explored in literatures, their thermodynamic properties are seldom discussed.

\begin{figure}
	\centering
	\includegraphics[width=0.41\textwidth]{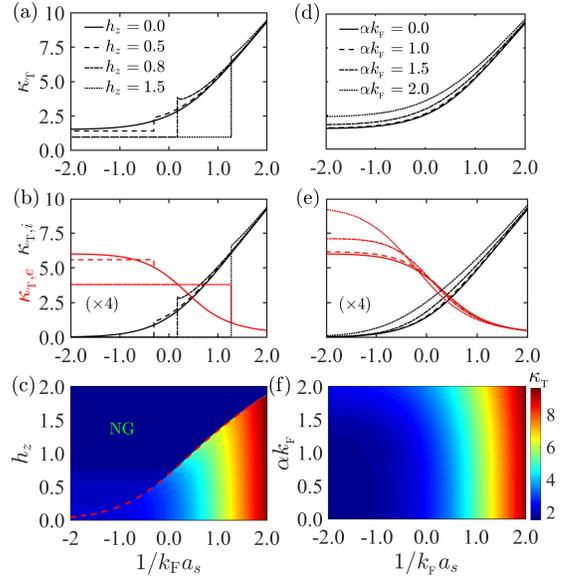}
	\caption{(Color online) Isothermal compressibility with only Zeeman field (left column) and SOC (right column). 
	(b)-(e) show the corresponding explicit and implicit compressibilities  $\kappa_{T,e/i}$ for these two cases.  
	(c)-(f) show the contour plot of isothermal compressibility as a function of Zeeman field (SOC) and scattering 
	length. In normal gas (NG), $\Delta = 0$. }
	\label{fig-fig1}
\end{figure}

In this work, we mainly focus on the effect of SOC and Zeeman field on isothermal compressibility, $\kappa_T$, which measures fluidity via \cite{lifshitz2013statistical, Pathria}
\begin{equation}
	\kappa_T = -{1\over V} \left({\partial V \over \partial P}\right)_{T, N}.
	\label{eq-def}
\end{equation}
Here the thermodynamic variables $P$, $V$, $T$ and $N$ correspond to pressure, volume, temperature and total number of particles, respectively. 
The minus sign in Eq. \ref{eq-def} is used to ensure $\kappa_T > 0$ for the stable phases. For an ideal gas, $\kappa_T = 1/P$; while in solid 
material, $\kappa_T = 1/B$, where $B$ is the corresponding bulk modulus \cite{kittel_introduction_2005}.
For ideal gas, this quantity is related to sound velocity via Newton-Laplace formula $c = \sqrt{\gamma V/N \kappa_T}$, 
where $\gamma$ is the isentropic expansion factor \cite{turns_thermodynamics:_2006}. According to this definition,
the more fluidity the system is, the larger this value will be. For this reason, this value was used in literatures to identify the 
boundaries between superfluid phases and insulating phases \cite{guo2013compressibility, wang2017topological, costa2018compressible, zhang2017ground, 
duarte_compressibility_2015}; as well as the boundary between normal phase and Bose-Einstein condensates (BEC). In BEC, $\kappa_T$ will divergent since the 
condensate does not contribute to pressure \cite{lifshitz2013statistical, Pathria, yukalov2007bose}. In experiments this quantity  has also been explored with both
fermions \cite{lee2012compressibility,ku2012revealing,duarte_compressibility_2015} and bosons \cite{poveda2015isothermal, castilho2016equation,gemelke_situ_2009}.

\begin{figure}
	\centering
	\includegraphics[width=0.41\textwidth]{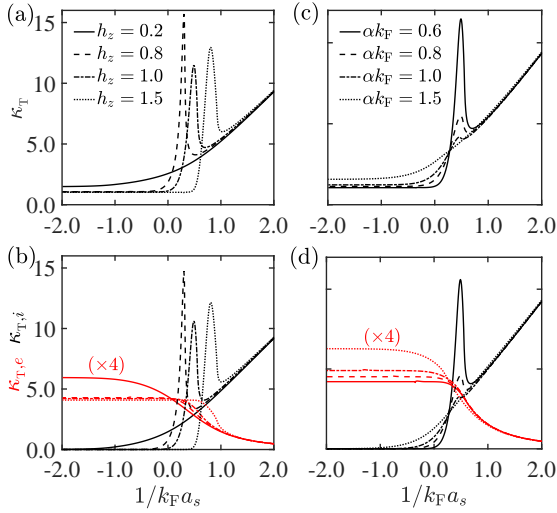}
	\caption{(Color online) (a) and (c) Isothermal compressibility with both Zeeman field and SOC. (b) and (d) show the corresponding explicit and implicit 
	compressibility. In (a)-(b), $\alpha k_F = 0.6$; and (c)-(d), $h_z = 1.0$.}
	\label{fig-fig2}
\end{figure}

We investigate the isothermal compressibility in the spin-orbit coupled degenerate Fermi gases. We formulate this quantity based on Gibbs-Duhem
equation \cite{perrot_z_1998}, and find that due to the implicit dependence of pairing strength on chemical potential and carrier density, this quantity 
can be divided into two parts: the explicit term related to chemical potential, and the implicit term related to order parameter. In the Bardeen-Cooper-Schrieffer
(BCS) limit, this value is determined by the density of state at the Fermi surface, while in the BEC limit it is determined by the scattering length. 
In the intermediate regime with both spin-orbit coupling and Zeeman field, we find a pronouncedly enhancement of compressibility in the gapless Weyl 
superfluid phase regime contributed from the implicit part. This kind of peak can be regarded as the remanent effect of phase separation (PS).  The similar 
features can also be found in an optical lattice away from half filling. 

\begin{figure}
	\centering
	\includegraphics[width=0.43\textwidth]{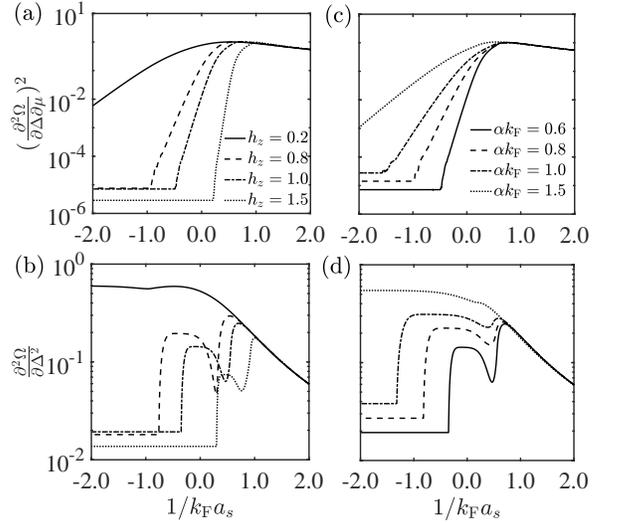}
	\caption{(Color online) The numerator and denominator of the implicit compressibility $\kappa_{T, i}$ as a function of scattering length 
	for different Zeeman field and SOC strengths. When $\Delta = 0$, $\kappa_{T,i} = 0$ accounts for the plateau regime in BCS limit. }
	\label{fig-fig3}
\end{figure}

{\it Theory}. We work in the grand canonical ensemble and the Gibbs thermodynamic potential $G = \sum_{\sigma} \mu_\sigma N_\sigma$ for a two-component
system with $\sigma = \uparrow, \downarrow$. In equilibrium, we have the following Gibbs-Duhem equation \cite{lifshitz2013statistical, Pathria},
\begin{equation}
	-SdT + VdP = \sum_{\sigma} N_{\sigma} d{\mu_{\sigma}},
	\label{eq-Gibbs}
\end{equation}
in which the chemical potential $\mu_{\uparrow} = \mu + h_z$ and $\mu_\downarrow = \mu-h_z$, with $h_z$ being the effective Zeeman field (see below), $S$ is the 
entropy of the whole system and $N = N_\uparrow + N_\downarrow$ is the total number of particle. In the case of fixed temperature, Eq. \ref{eq-Gibbs} 
establishes a direct connection between pressure and carrier density, $P = P(T, n_{\uparrow}, n_{\downarrow})$, with $n_\sigma = N_\sigma/V$, since pressure 
is an intensive quantity. Then from the differential chain-rule, we obtain 
\begin{equation}
    {1 \over \kappa_T} = -V \left({\partial P \over \partial n_{\uparrow}}\right)_{T, n_{\downarrow}}
    \left({\partial n_{\uparrow} \over \partial V}\right)_{T, n_{\downarrow}} 
    -\left(\uparrow \leftrightarrow \downarrow\right).
\end{equation}
From the Maxwell relation, ${\partial P \over \partial N_{\sigma}} =- {\partial \mu_\sigma\over \partial V}$, we have
\begin{equation}
    {1 \over \kappa_T} = - \sum_{\sigma} N_{\sigma} \left({\partial \mu_\sigma \over \partial V}\right)_T.
\end{equation}
One should notice that the chemical potential is also an intensive quantity, that is, $\mu_\sigma = \mu(n_{\sigma},T)$. 
Let us define compressibility matrix as $\kappa_{ij}^{-1} = n_i n_j {\partial \mu_{i} \over \partial n_j}$, then the isothermal compressibility can 
be written as,
\begin{equation}
    {1 \over \kappa_T} = \sum_{i,j \in\uparrow, \downarrow} {1 \over \kappa_{ij}}.
    \label{eq-kappaT}
\end{equation}
This relation can be generalized to arbitrary number of components. In the limiting case when $n_i$ is independent of $\mu_j$ for $i \ne j$, one finds 
$\kappa_{ij}^{-1} = 0$. Otherwise, $\kappa_{ij}^{-1} \ne 0$, thus the right hand side is always well-defined. Moreover, in the limiting case when $\mu_\sigma = \mu$, 
the above compressibility is reduced to $\kappa_T = {1 \over n^2} \left({\partial n \over \partial \mu}\right)_T$,
which was widely used in literatures \cite{guo2013compressibility, wang2017topological, costa2018compressible, zhang2017ground, duarte2015compressibility}.
Finally, let us stress that even at zero temperature, this quantity is nonzero.

In following we employ the above theory to understand the fluidity of the spin-orbit coupled superfluids in free space. The single particle term can be written as 
$H_0 = \sum_{{\bf k} \sigma \sigma^\prime} c_{{\bf k}\sigma}^\dagger [\xi_{\bf k} + \alpha (k_y \sigma_x - k_x \sigma_y) + h_z \sigma_z]_{\sigma\sigma^\prime} c_{{\bf k}\sigma^\prime}$,
where $\xi_{\bf k}  = \epsilon_{\bf k} - \mu = {k^2 \over 2m}-\mu$, ${\bf k} = (k_x, k_y, k_z)$, $\alpha$ is the SOC coefficient, and $h_z$ is the corresponding Zeeman field 
between the two components and $\sigma_{x,y,z}$ are Pauli matrices. In the presence of $s$-wave interaction between the two species, one can define a uniform 
pairing order $\Delta = {g \over V} \sum_{\bf k} \langle c_{-{\bf k}\uparrow} c_{{\bf k}\downarrow}\rangle$, where $g$ is the scattering strength. 
Let us define the thermodynamic potential $\Omega$ through the partition function $Z = e^{-\beta \Omega} = 
\text{Tr}(e^{-\beta H})$, with $\beta = 1/k_B T$, then 
\begin{equation}
    \Omega = \sum_{\bf k} \xi_k - {1\over \beta} \sum_{{\bf k}\lambda} \ln \left[2\cosh(\beta E_{{\bf k}}^\lambda/2)\right] - {V|\Delta|^2 \over g},
	\label{eq-omega}
\end{equation}
where $E_{\bf k}^\lambda = \sqrt{\abs{\gamma_{\bf k}}^2 + \xi_{\bf k}^2 + h_z^2 + |\Delta|^2 + 2\lambda E_0}$ is the excitation spectra, 
$E_0 = \sqrt{h_z^2(\xi_{\bf k}^2 + |\Delta|^2) + |\gamma_{\bf k}|^2 \xi_{\bf k}^2}$, 
$\abs{\gamma_{\bf k}}^2=\alpha^2(k_x^2+k_y^2)$ and $\lambda = \pm 1$. The corresponding carrier  
density and order parameter  are determined by $n_\sigma = -{1\over V} {\partial \Omega \over \partial \mu_\sigma}$, ${\partial \Omega \over \partial \Delta} = 0$. 
During regularization, in Eq. \ref{eq-omega}, we have used ${1\over g} = {m \over 4\pi a_s} - {1\over V} \sum_{\bf k} {1 \over {k^2/m}}$, with $a_s$ being the 
scattering length. For more details, please see Ref. \cite{gong2011bcs}. 

The important point is that, though $\Delta$ is an important quantity to characterize the interaction between the particles. It is not a thermodynamic 
variable. Thus to compute $\kappa_T$ in Eq. \ref{eq-kappaT}, we may explicitly ($e$) take derivative of carrier density with respect to chemical potential, or 
implicitly ($i$) take derivative of carrier density with respect to order parameter $\Delta$, as following,
\begin{equation}
	\left({\partial n_i \over \partial \mu_j}\right)_T =
        \left({\partial n_i \over \partial \mu_j}\right)_{T, i}
        + \left({\partial n_i \over \partial \mu_j}\right)_{T, e},
\end{equation}
where $({\partial n_i \over \partial \mu_j})_{T, i} = -({\partial^2 \Omega \over \partial \mu^2})_{T, \Delta}$. In the second term, 
\begin{equation}
	\left({\partial n_i \over \partial \mu_j}\right)_{T, i} =
        \left({\partial n_i \over \partial \Delta}\right)_{T,e} 
        \left({\partial \Delta \over \partial \mu_j}\right)_{T, i}.
\end{equation}
Let's define $f = {\partial \Omega \over \partial \Delta} = 0$,
then using $\dd f = {\partial f \over \partial \mu} \dd\mu + {\partial f \over \partial \Delta} \dd\Delta = 0$ we find
$(\frac{\partial\Delta}{\partial\mu})_{T,i} =-(\frac{\partial^{2}\Omega}{\partial\mu\partial\Delta}/\frac{\partial^{2}\Omega}{\partial^{2}\Delta})$,
thus $({\partial n \over \partial \mu})_{T, i} = 
\left(\frac{\partial^{2}\Omega}{\partial\Delta\partial\mu}\right)_{T,e}^{2}/\frac{\partial^{2}\Omega}{\partial^{2}\Delta}$.
Collecting these results together yields
\begin{equation}
	\kappa_T = \kappa_{T,e} +  \kappa_{T,i},
	\label{kt-tot}
\end{equation}
where their expressions are presented below:
\begin{equation}
    \kappa_{T,e}= 
    {{\displaystyle\sum\limits_{\mathbf{k},\lambda}}
		\left(  Y_{{\bf k}}^{\lambda}-X_{ {\bf k}}^{\lambda}\right)
        \left(
			\frac{\xi_{\mathbf{k}} Q_{ {\bf k}}^{\lambda}} {E_{\mathbf{k}}^\lambda}
        \right)^2
		+X_{ {\bf k}}^{\lambda}
        \left(  
			Q_{ {\bf k}}^{\lambda}-\lambda\frac{ \xi_{\mathbf{k}}^2P_{\bf k}^2 }{E_{0}^{3}}
        \right)
	\over 2 n^{2}},
\label{eq-kte}
\end{equation}
and
\begin{equation}
\kappa_{T,i} = 
\dfrac{ 
\left\{ 
{\displaystyle \sum\limits_{\mathbf{k},\lambda}}
\xi_{\mathbf{k}}
\left[
(X_{\mathbf{k}}^\lambda - Y_{\mathbf{k}}^\lambda) 
\frac{Q_{\mathbf{k}}^\lambda S_{\mathbf{k}}^\lambda}{{E_{\mathbf{k}}^\lambda}^2} 
+ \lambda X_{\mathbf{k}}^\lambda \frac{h_z^2 P_{\mathbf{k}}}{E_0^3}
\right]
\right\}^2 }
{2n^2 
{\displaystyle \sum\limits_{\mathbf{k},\lambda}}
\left[
(X_{\mathbf{k}}^\lambda - Y_{\mathbf{k}}^\lambda)
\left(\frac{S_{\mathbf{k}}^\lambda}{E_{\mathbf{k}}^\lambda} \right)^2 
- \lambda X_{\mathbf{k}}^\lambda 
  \frac{h_z^4}{E_0^3}
  \right]}.
  \label{eq-kti}
\end{equation}
Here $X_{{\bf k}}^\lambda = \tanh\left(\frac{\beta E_{\bf k}^\lambda}{2}\right)/E_{\mathbf{k}}^\lambda$, $Y_{{\bf k}}^\lambda = \beta (1-\tanh^2(\frac{\beta E_{\mathbf{k}}^\lambda}{2}))/2$ 
(with $X_{{\bf k}}^\lambda \ge Y_{{\bf k}}^\lambda$ for any $\beta$), $P_{\bf k}=h_z^2+\abs{\gamma_{\bf k}}^2$, $Q_{ {\bf k}}^\lambda=1+\lambda P_{\bf k}/E_0$ and $S_{ {\bf k}}^\lambda=1+\lambda h_z^2/E_0$. 

Eq. \ref{eq-kte} and \ref{eq-kti} are two major results we have obtained in this work. Before presenting our numerical results, let us 
discuss the implementations of these results in some limiting cases. 
(i) Without interaction, $\Delta = 0$, the implicit term $\kappa_{T, i} = 0$,
and the total carry density $n = {1\over V} \sum_{{\bf k}\lambda} n_{{\bf k}\lambda}$,
thus $\kappa_T = {\beta \over 4n^2 V } \sum_{{\bf k}\lambda} \left[1 - \tanh^2(\beta E_{{\bf k}\lambda}/2)\right]$.
Obviously, at zero temperature, $\kappa_T = {1 \over 2n^2 V} \rho(\mu) \propto \rho(\mu)$, where $\rho$ denotes
for density of state at the Fermi surface. Thus in insulating phase with $\rho(\mu) = 0$, $\kappa_T = 0$. (ii) Without SOC and Zeeman field, we have $E_0 = 0$ and 
$E_{{\bf k} }^\lambda = E_{\bf k}$, then $\kappa_{T,i} = {1\over n^2 V} (\sum_{\bf k} \xi_{\bf k}/E_{\bf k}^3)^2/(\sum_{\bf k} 1/E_{\bf k}^3)$, and $\kappa_{T, e} = {1\over n^2 V} \sum_{\bf k} 
(1/E_{\bf k} - \xi_{\bf k}^2/E_{\bf k}^3)$, where $E_{\bf k}=\sqrt{\xi_\textbf{k}^2+|\Delta|^2}$. Obviously, both the implicit part and explicit part are positive value. This case can also be 
computed exactly using ${1\over V} \sum_{\bf k} {\xi_{\bf k} \over E_{\bf k}^3} = {K(x) \over 4\pi^2 (\Delta^2 + \mu^2)^{1/4}}$ and ${1\over V} \sum_{\bf k} {1 \over E_{\bf k}^3} = {2\sqrt{\mu^2 + \Delta^2}E(x) 
+ (\mu - \sqrt{\mu^2 + \Delta^2})K(x) \over 4\pi^2 (\Delta^2 + \mu^2)^{1/4}}$, where $E(x)$ and $K(x)$, with $x = {1\over 2}(1 + {\mu \over \sqrt{\mu^2 + \Delta^2}})$, 
are the second kind incomplete and the first kind complete elliptic integrals, respectively. With these expressions,
we find that the compressibility $\kappa_{T, e} \gg \kappa_{T, i}$ and  $\kappa_{T, i} \sim 0$ in the BCS limit, while in the 
BEC limit, $\kappa_{T, e}  \sim 0$, while $\kappa_{T, i} \propto \sqrt{-\mu}$. Notice that in the BEC limit,
$\mu \propto -{1 \over (k_Fa_s)^2}$, thus we find $\kappa_{T, i} \propto {1 \over k_F a_s}$ 
(see numerical results in Fig. \ref{fig-fig1}). This result also explains the linear behavior of compressibilities  in the BEC limit with 
both SOC and Zeeman field in Fig. \ref{fig-fig2}. (iii) The case with only SOC was investigated in Ref. \cite{han2012evolution}, and our expression can be reduced to
the results in Ref. \cite{han2012evolution} by letting $h_z = 0$. The interesting point is that in the presence of
both SOC and Zeeman field, the excitation spectra may become gapless in the Weyl superfluids regime.
The expression for $\kappa_{T, i}$ is no longer always larger than zero, giving rise to PS phase. We have utilized this feature.

\begin{figure}
	\centering
	\includegraphics[width=0.45\textwidth]{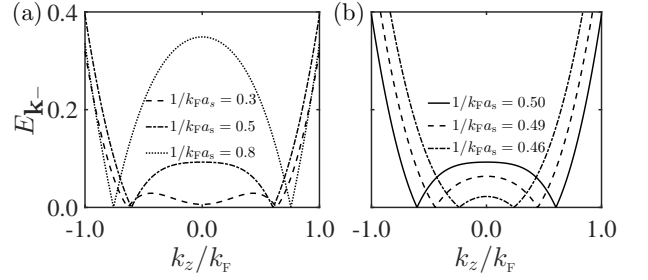}
	\caption{(Color online) Band structure of $E_{{\bf k}-}$ at the isothermal compressibility peaks in Fig. \ref{fig-fig2}. 
	We have setted $k_x = 0$ and $k_y = 0$, which is the case for gapless Weyl points.}
	\label{fig-fig4}
\end{figure}

{\it Numerical results}. We determine the value of $\mu$ and $\Delta$ self-consistently at zero temperature \cite{de2008fermions, de1993crossover}. 
The Fermi momentum $k_F = \sqrt[3]{3\pi^2n}$ and Fermi energy $E_F = k_F^2/(2m)$ serve as basic scales for momentum and energy, respectively. 
We first discuss the role of SOC and Zeeman field individually in Fig. \ref{fig-fig1}. In strong Zeeman field and in the BCS limit, the pairing is completely destroyed when $|h_z| > |\Delta|$, 
thus we have a normal gas (NG) phase. In both cases we find that $\kappa_T$ in the BEC limit is much larger than that in the BCS side. Moreover, since the Zeeman field plays the role of 
reducing the density of state at the Fermi surface, while SOC plays the opposite role, we find the same trend for the isothermal compressibility.
We also find that in the BCS limit, $\kappa_{T, e} \gg \kappa_{T, i} \sim 0$, while in the BEC side, $\kappa_{T, i} \gg \kappa_{T, e} \sim 0$, as expected from our theoretical analysis.

The physics is completely changed in the presence of both terms; see Fig. \ref{fig-fig2}. We find a pronouncedly enhancement of isothermal compressibility, by one order of magnitude, 
in some proper parameter regimes. This enhancement compressibility is more likely to be found in regime with relative larger Zeeman field and weaker SOC. Especially, we find that the peak position depends
more strongly on the Zeeman field. In Fig. \ref{fig-fig2}b and d, we find that this peak arises from the implicit part, while the explicit part always shows a smooth behavior. To further pin down the reason for 
this anomalous peak, we plot the numerator and denominator of $\kappa_{T, i}$ in Fig. \ref{fig-fig3}. The numerator is always a smooth function of scattering length. However, the denominator 
exhibits some peculiar behavior with increasing of scattering length from BCS side to the BEC side. The dip in the denominator accounts for the anomalous behavior of compressibility. We find that in 
the fully gapped regime, the denominator is always very large; it can take a minimal value only in the gapless regimes, which can be realized either with $\Delta \sim 0$ (BCS limit)
or Weyl superfluid phase regime. We illustrate this physics in Fig. \ref{fig-fig4} by plotting the band structure of the corresponding peaks in Fig. \ref{fig-fig2}, which are 
always gapless in the Weyl superfluids. With the increasing of scattering 
length in the BEC side, these Weyl points are destroyed, and the superfluid enter the fully gapped phase,
in which the compressibility becomes extremely large due to condensation, and we have $\kappa_T \propto {1 \over k_F a_s}$.

\begin{figure}
	\centering
	\includegraphics[width=0.45\textwidth]{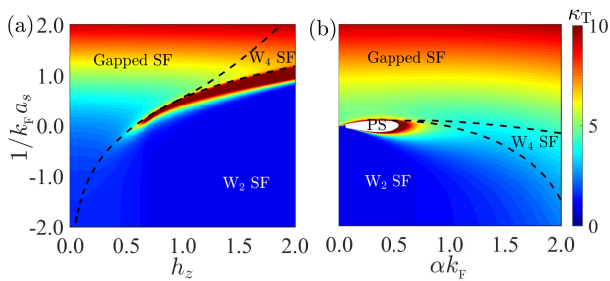}
	\caption{(Color online) Contour plot of the isothermal compressibility and phase diagram. The white regime corresponds to the unstable PS phase in the presence of imbalance $h_z$
	and SOC, identified by $\kappa_{T} < 0$. In (a) we have used $\alpha k_F=0.6$ and in (b) $h_z=1.0$. W$_{2/4}$ SF denote topological Weyl supefluids with two and
	four Weyl points, respectively.}
	\label{fig-fig5}
\end{figure}

We plot the compressibility and phase diagram as a function of Zeeman field and SOC in Fig. \ref{fig-fig5}. There is a small regime for PS, which is determined by 
$\kappa_{T} < 0$. In this case the Helmholtz free energy $F = \Omega + \sum_\sigma \mu_\sigma  N_\sigma $ shows two local minimals in space by $\mu$ and $\Delta$. The SOC can effectively 
suppress this PS effect. Near this regime, by tuning of Zeeman field, we can find a dramatic enhancement of isothermal compressibility in Fig. \ref{fig-fig5}a near the boundary between the
fully gapped superfluid and the Weyl superfluids with two Weyl points (W$_2$ SF). Th peak position in the anomalous regime depends strongly on the Zeeman field. Thus we may regard this 
anomalous behavior as the remanent effect of PS, which happens near the boundary between gapped phase and gapless Weyl phase.

Finally we have also examined the same quantity in the lattice model, in which similar anomalous behavior have also been identified. In the optical lattice, the filling factor and particle-hole
symmetry about half filling become two important controlling parameters in experiments. We find that near half filling, the implicit compressibility is greatly suppressed, while away from 
this regime, this kind of anomalous compressibility can always be found. These results will be discussed elsewhere. In experiments, the enhancement of compressibility by one order of magnitude 
is arresting and can be revealed from the anomalous behavior of sound velocity from Newton-Laplace equation $c\propto \kappa_T^{-1/2}$, or the density fluctuation via 
$\langle \delta n^2\rangle \sim \kappa_T k_B T$ \cite{lifshitz2013statistical, Pathria}.

To conclude, we present a general theory to study the isothermal compressibility in the superfluids with both SOC and Zeeman field from the Gibbs-Duhem equation. These two terms can modify 
the band structure and possible pairings, 
thus dramatically influences its isothermal compressibility. We find that in the BCS limit the compressibility is determined by the density of state at the Fermi surface,
while in the BEC limit, it is determined by the scattering length. Between these two regimes,  we predicted a pronouncedly enhancement of isothermal compressibility in the gapless Weyl phases. 
The peak mainly comes from the implicit contribution of the order parameter. This kind of behavior can be found in both free space and optical lattice models. The foundation in this work pave 
the way for exploring other thermodynamic properties in spin-orbit coupled ultracold atoms.

\textit{Acknowledgements.} The authors thank H. G. Luo for the great help in carrying out numerical calculation.
M.G. is supported by the National Youth Thousand Talents Program (No. KJ2030000001), the USTC start-up funding (No. KY2030000053),
the national natural science foundation (NSFC) under grant No. 11774328). M. G. and
G. G are supported by The National Key Research and Development Program of China (No. 2016YFA0301700).
Y. Z. is supported by the NSFC under Grants NO. 11674201 and NO. 11474189.

\bibliographystyle{apsrev4-1}
\bibliography{ref}

\end{document}